\documentclass[twocolumn,amsmath,amssymb]{revtex4}
\usepackage{graphicx}
\usepackage{dcolumn}
\usepackage{bm}
\begin{document}

\title{Aspects of Anisotropic Fractional Quantum Hall Effect in Phosphorene}
\author{Areg Ghazaryan and Tapash Chakraborty\footnote{Tapash.Chakraborty@umanitoba.ca}}
\affiliation{Department of Physics and Astronomy,
University of Manitoba, Winnipeg, Canada R3T 2N2}

\date{\today}
\begin{abstract}
We have analyzed the effects of the anisotropic energy bands of phosphorene on 
magnetoroton branches for electrons and holes in the two Landau levels close to 
the band edges. We have found that the fractional quantum Hall effect gap in the 
lowest (highest) Landau level in conduction (valence) band is slightly larger 
than that for conventional semiconductor systems and therefore experimentally 
observable. We also found that the magnetoroton mode for both electrons and holes 
consists of two branches with two minima due to the anisotropy. Additionally, 
we show that due to the anisotropy, there is a second mode with positive dispersion,
well separated from the magnetoroton mode for small wave vectors. These novel features
of the collective mode can be observed in resonant inelastic light scattering experiments.

\end{abstract}

\maketitle
With the advent of graphene \cite{graphene_book,abergeletal} the two dimensional 
(2D) materials have gained renewed attention during the last decade because of 
their remarkable electronic, optical and mechanical properties and for potential 
device applications. Practical application of graphene in electronic devices 
is however limited due to its zero band gap. Although there are external means to 
create band gap in graphene, such as application of a bias voltage to bilayer graphene, 
in the past few years intense search has been undertaken to find other 
2D materials, which in addition to exhibiting mobilities close to graphene, also 
posses a considerable band gap in their normal state. There were several Dirac 
materials, such as silicene \cite{silicene}, germanene \cite{germanene}, $\mathrm{MoS_2}$ 
and other group-VI transition-metal dichalcogenides \cite{MoS2} that have been lately explored 
and promising results were observed.

In the past year another material which has gained popularity in this context, is the 2D version 
of black phosphorus (BP) \cite{LiNature,LiuAcsNano,XiaNature,Kou,LiuReview}, the single 
layer of BP known as phosphorene. BP is the most stable allotrope of Phosphorus at room 
temperature and pressure. Few-layer BP can be obtained through mechanical exfoliation 
method akin to graphene. However, unlike graphene the BP layers are not perfectly 
flat. Due to the $sp^3$ hybridization of $3s$ and $3p$ atomic orbitals the layers of BP form 
a puckered surface. The band gap of BP depends 
on the number of layers of the sample, and it ranges from 0.3 eV in the bulk case at the Z point 
in the Brillouin zone to 1.5 - 2 eV for few layer and monolayer case at the $\Gamma$ 
point. The mobilities of the few-layer BP obtained so far lie in the range of 1000 - 5000 
$\mathrm{cm^2V^{-1}s^{-1}}$ \cite{LiNature, LiHall1, LiHall2}. Considerable progress has
been made in the study of the electronic and optical properties of this material both 
experimentally \cite{LiNature,LiuAcsNano,XiaNature,Koenig,LiHall1,LiHall2} and theoretically 
\cite{Rodin,Low,Jiang,Rudenko,Yuan,Pereira,Zhou,LiuNano,Ezawa,Tahir,Cakir}. Higher mobilities 
were achieved by sandwiching the BP in hexagonal boron nitride flakes and placing on top of 
the graphite back gate that has enabled observation of the integer quantum Hall effect 
\cite{LiHall2}. Clearly the next step in the experimental progress is to reach the regime 
where the fractional quantum Hall effect (FQHE) \cite{FQHEBook} can be observed.

The FQHE in Dirac materials shows many interesting features not found in conventional 
semiconductor quantum heterostructures. The FQHE states in monolayer and bilayer graphene were 
investigated theoretically \cite{FQHE_chapter,ApalkovMonolayer,ApalkovBilayer1,ApalkovBilayer2} and 
experimentally \cite{FQHEGraphene,FQHEBilayer}. It was shown that while the FQHE states in 
graphene in the $n=0$ Landau level (LL) are completely identical to the case of conventional 
semiconductor systems, graphene exhibits more robust FQHE states in the $n=1$ LL with a bigger 
gap than in the $n=0$ LL \cite{ApalkovMonolayer}. That behavior is not present in 
conventional semiconductor systems. In bilayer graphene the application of a bias voltage results 
in some LLs the observation of the phase transition between several incompressible FQHE and 
compressible phases \cite{ApalkovBilayer1,FQHEBilayer}. The FQHE in silicene and germanene 
indicated that because of the strong spin-orbit interaction present in these materials as
compared to graphene, the electron-electron interaction and the FQHE gap are significantly modified 
\cite{ApalkovBuckled}. The puckered structure of phosphorene exhibits a lower symmetry than graphene. 
This results in anisotropic energy spectra and other physical characteristics of phosphorene, both 
in momentum and real space in the 2D plane \cite{Kou,LiuReview}. Therefore it is desirable to explore 
how this anisotropy of phosphorene manifests itself in the FQHE states.

In this work we consider the FQHE states in phosphorene for filling factor $\nu=1/3$ in the
conduction band and $\nu=2/3$ in the valence band in two LLs closest to the band edges. We analyze
the energy spectra and the magnetoroton modes of the system with finite number of electrons using 
the exact diagonalization scheme. We observe the FQHE gap in the lowest (highest) LL in the
conduction (valence) band, with the gap being slightly higher than in conventional semiconductor 
systems. This indicates that the FQHE should be observable in phosphorene. We also show that both the 
gap of the FQHE and the magnetoroton mode are similar for the lowest LL in the conduction band and 
highest LL in the valence band. The anisotropy of phosphorene manifests itself in two different 
ways. We find that the magnetoroton mode in phosphorene (for electrons and holes) is split 
into two branches with two minima due to the anisotropy. We also find that the anisotropy
causes a second mode with positive dispersion to appear that is well separated from the 
magnetoroton mode for small wave vectors and is therefore experimentally more accessible than for 
isotropic systems. For the FQHE states in the second LLs of the conduction and valence band of 
phosphorene, the FQHE gap is an order of magnitude smaller than for the lowest LLs and approaches 
a soft mode, with possible incompressible to compressible phase transition.         

In what follows we consider a finite-size system of electrons in 2D phosphorene in the toroidal 
geometry using the exact diagonalization procedure \cite{FQHEBook}. The many-body Hamiltonian of 
the system is
\begin{equation}
\label{MBHam}
{\cal H} = \sum_i^{N^{}_e}{\cal H}^i_\mathrm{P} + \frac12\sum_{i\neq j}^{N^{}_e}V^{}_{ij},
\end{equation}
where ${\cal H}_\mathrm{P}$ is the single-particle Hamiltonian in phosphorene and the second term 
is the Coulomb interaction. Since we are interested in the FQHE in the low lying LL, which are 
situated close to the conduction and the valence band edges, we employ the continuum approximation 
of the single-particle Hamiltonian derived from the microscopic tight-binding approach \cite{Pereira}. 
In all our calculations we take the $x$ axis along the armchair direction and the
$y$ axis along the zigzag direction. Then the single-particle Hamiltonian in single-layer phosphorene 
in the continuum approximation has the form \cite{Pereira}
\begin{align}
\label{SPHam}
{\cal H}^{}_\mathrm{P}(\mathbf{k}) &=\left(\begin{array}{cc}
 H^{}_0(\mathbf{k}) & H^{}_1(\mathbf{k}) \\
H^{*}_1(\mathbf{k}) & H^{}_0(\mathbf{k}) \end{array}\right),\\
H^{}_0(\mathbf{k})&=u^{}_0+\eta^{}_xk_x^2+\eta^{}_yk_y^2, \nonumber \\
H^{}_1(\mathbf{k})&=\delta+\gamma^{}_xk_x^2+\gamma_yk_y^2+i\chi k^{}_x, \nonumber
\end{align}
where $\mathbf{k}$ is the wave vector and all parameters in $H^{}_0(\mathbf{k})$ and 
$H^{}_1(\mathbf{k})$ are related to five hopping amplitudes in the tight-binding model 
\cite{Pereira}: $u^{}_0=-420\,\mathrm{meV}$, $\eta^{}_x=10.1\,\mathrm{meV\cdot 
nm^2}$, $\eta^{}_y=5.8\,\mathrm{meV\cdot nm^2}$, $\delta=760\,\mathrm{meV}$, $\gamma^{}_x=38.3\,
\mathrm{meV\cdot nm^2}$, $\gamma^{}_y=39.3\,\mathrm{meV\cdot nm^2}$, $\chi=525\,\mathrm{meV\cdot 
nm}$. For these parameters the band edges of the conduction and the valence band are located 
at the energies $E^{}_c=340\,\mathrm{meV}$ and $E^{}_v=-1180\,\mathrm{meV}$, with energy gap 
$E^{}_g=1520\,\mathrm{meV}$. In a magnetic field, we first write $\mathbf{k}=-i\nabla+e\mathbf{A}/
\hbar c$, where the vector potential is in the Landau gauge $\mathbf{A}=Bx\mathbf{\hat{y}}$. 
For all the FQHE states considered here the many-body ground state is fully spin polarized. We 
henceforth disregard both the electron spin and the Zeeman energy. For the next step we cast the 
Hamiltonian (\ref{SPHam}) into the second quantized form using the ladder operator
$a =\frac1{\sqrt{\hbar\omega^{}_c}}\left[\sqrt{\eta^{}_x}k^{}_x-i\sqrt{\eta^{}_y}k^{}_y\right],$ 
where $\hbar\omega^{}_c=2\sqrt{\eta^{}_x\eta^{}_y}/\ell^2_0$ and $\ell_0=\sqrt{\hbar c/eB}$ is 
the magnetic length. It is easy to verify that $[a,a^\dagger]=1$. Then the components of the 
Hamiltonian (\ref{SPHam}) are
\begin{align*}
H^{}_0 &=u_0+\hbar\omega^{}_c\left(a^\dagger a+\frac12\right), \nonumber \\
H^{}_1 &=\delta+\hbar\omega^{}_c\left(s\left(a^\dagger a+\frac12\right)+\frac14(s-t)\left(
a^\dagger-a\right)^2\right) \\
&+i\hbar\omega^{}_c\tilde{\chi}\left(a^\dagger+a\right),
\end{align*}
where $s=\gamma^{}_x/\eta^{}_x$, $t=\gamma^{}_y/\eta^{}_y$, $\tilde{\chi}=\chi/2\sqrt{\eta^{}_x
\hbar\omega^{}_c}$. We search for the eigenvalues of the Hamiltonian (\ref{SPHam}) in the form
\begin{equation}
\label{SPWaveFunc}
\psi(x,y)=\left(\begin{array}{c} \sum_{n=0}^{n^{}_\mathrm{max}} A^{}_n\phi^{}_{n,k^{}_y}(x,y) \\
\sum_{n=0}^{n^{}_\mathrm{max}} B^{}_n\phi^{}_{n,k^{}_y}(x,y)
\end{array}\right),
\end{equation}
where $A^{}_n$ and $B{}_n$ are constants and $\phi_{n,k^{}_y}(x,y)$ are the
usual harmonic oscillator wave functions 
\begin{align}
\label{HOWave}
\phi_{n,k^{}_y}(x,y)=\frac{\alpha^{\frac14}}{\sqrt{2^nn!L^{}_y\sqrt{\pi}}} & e^{ik^{}_y y}e^{-\frac{\alpha
\left(x+k^{}_y\ell^{2}_0\right)^2}{2}} \nonumber \\
& \times H^{}_n\left(\sqrt{\alpha}\left(x+k^{}_y\ell^2_0\right)\right),
\end{align}
where $\alpha=\hbar\omega^{}_c/2\eta^{}_x$, $L^{}_y$ is the sample size in the $y$ direction and 
$k^{}_y$ is the wave vector in the $y$ direction, which is conserved in the chosen gauge. 
Diagonalizing the Hamiltonian matrix obtained with the wave function (\ref{SPWaveFunc}) 
we determine the eigenvalues of the Hamiltonian (\ref{SPHam}), which are the LLs of phosphorene and 
the appropriate wave functions by determining the constants $A^{}_n$ and $B{}_n$.

In order to consider the many-body states we proceed in the same way as for conventional 
semiconductor systems \cite{FQHEBook,Haldane}. We consider the many-body states in the system 
in a toroidal geometry, i.e.,  we apply the periodic boundary conditions (PBC) on 
both directions of the system \cite{FQHEBook,Haldane,PBCpapers} and cast the wave functions 
(\ref{SPWaveFunc}) and (\ref{HOWave}) into the PBC preserving form. This process naturally defines 
the magnetic translation unit cell, which we take rectangular in our calculations with the sides 
$L^{}_x$ and $L^{}_y$, $L^{}_xL^{}_y=2\pi\ell^2_0N^{}_s$, where $N^{}_s$ is the number 
of magnetic flux quanta passing through the unit cell and is an integer number. $N^{}_s$ also 
characterizes the degeneracy of each LL. We take $N^{}_e=pN$ and $N^{}_s=qN$, where $N^{}_e$ is the 
number of electrons, $N$ is the greatest common divisor of $N^{}_e$ and $N^{}_s$, $p$ and $q$ are 
integers. The filling factor in a LL is defined as $\nu=p/q$. We then construct the
many-body states from the appropriate single-particle states and calculate the Hamiltonian matrix 
of the form (\ref{MBHam}) with this many-body basis states. We make use of the many-body translational
symmetry to reduce the size of the Hilbert space \cite{FQHEBook,Haldane}: We define the
relative translation operator $T^{\mathrm{R}}_i(\mathbf{a})$ and note that the translations  
$T^{\mathrm{R}}_i(p\mathbf{L}_{mn})$ preserve the Hilbert space and commute with the Hamiltonian 
(\ref{MBHam}) and with each other. Here $\mathbf{L}_{mn}=m\mathbf{L}_x+n\mathbf{L}_y$ is the magnetic 
translations lattice vector, $\mathbf{L}_x$ and $\mathbf{L}_y$ are the unit vectors of magnetic 
translations unit cell, $m$ and $n$ are integers. Therefore we use these relative translation operator 
($T^{\mathrm{R}}_i(p\mathbf{L}_{mn})$ eigenstates to bring the complete Hamiltonian 
matrix into the block-diagonal form and also to characterize each of the many-body states with its 
appropriate relative momentum $\mathbf{k}_\mathrm{R}$.

\begin{figure}
\includegraphics[width=7.0cm]{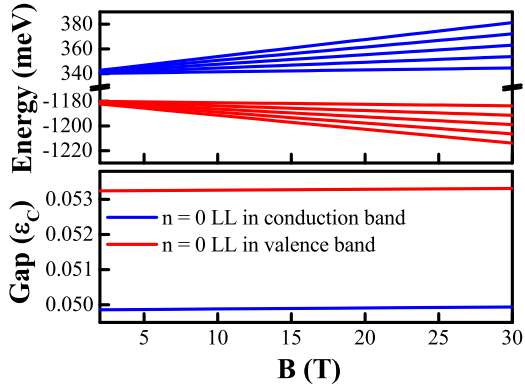}
\caption{\label{fig:LLGap_depB} (a) Magnetic field dependence of few low lying LLs in
phosphorene, both in the conduction (blue) and the valence (red) band. (b) Magnetic field dependence 
of the $\nu=1/3$ gap in the conduction band and the $\nu=2/3$ gap in the valence band lowest LLs of 
phosphorene. The finite-size system considered for FQHE gap has six electrons.}\end{figure}    

In Fig.~\ref{fig:LLGap_depB} (a) the magnetic field dependence of the few low lying LLs both in 
the conduction band and in the valence band of phosphorene is shown. The dependence is almost linear and 
is similar to the case of conventional semiconductors. In fact as was shown previously by several 
authors \cite{Jiang,Yuan,Pereira,Zhou} in the low energy regime the energy bands of the conduction 
and the valence band can be approximated by parabolic dispersion with anisotropic masses in the
$x$ and $y$ directions. Therefore for low-lying LLs and for magnetic fields up to $\sim$25 T the 
dependence of the LLs on the magnetic field is linear. For this reason, in considering the FQHE 
states in the low-lying LLs we can disregard the LL coupling (the possibility for electrons to occupy 
states in higher LL). In case of systems with parabolic dispersion the LL mixing parameter $\kappa$, the 
ratio between the Coulomb interaction energy and the cyclotron energy has the $\propto1/\sqrt{B}$ 
dependence  \cite{Peterson}. Therefore for high magnetic fields $\kappa<1$ and LL mixing 
can be safely ignored in our work.

In Fig.~\ref{fig:LLGap_depB} (b) the magnetic field dependence of the FQHE magnetoroton gap is 
presented for the lowest LL in the conduction and in highest LL in the valence band. Due to the 
particle-hole symmetry the filing factor $\nu=1/3$ of electrons in the valence band corresponds to 
$\nu=2/3$ filling for holes. The results are presented for the finite-size system with
$N^{}_e=6$. The aspect ratio of the magnetic unit cell $\lambda=L^{}_x/L^{}_y$ is $\lambda=N^{}_e/4$ 
in all calculations of the FQHE states. As the gap is presented in units of $\epsilon^{}_\mathrm{C}
=e^2/\epsilon\ell^{}_0$, where $\epsilon=10.2$ is the dielectric constant, this dependence is due 
to the mixing of each LL in phosphorene in terms of the harmonic oscillator wave functions 
(\ref{HOWave}). In conventional semiconductors the FQHE gap in units of $\epsilon^{}_\mathrm{C}$ 
does not depend on the magnetic field and for a similar system size the gap is 0.041. Therefore 
as seen in Fig.~\ref{fig:LLGap_depB} (b) the FQHE gaps in phosphorene are slightly larger than in 
conventional systems. The FQHE gap for the LL in the valence band is in fact larger than the gap in 
the conduction band, but overall all these gaps fall within the same energy range. The dependence 
of the FQHE gaps on the magnetic field is linear, albeit barely discernable, which again indicates that up 
to 25 T the non-trivial character of the LLs in phosphorene compared to conventional systems with 
parabolic dispersion is less important. The main features of phosphorene that can distinguish it 
from conventional semiconductor systems in the context of FQHE are the anisotropic nature of 
its bands, which we address next.       

\begin{figure}
\includegraphics[width=7.5cm]{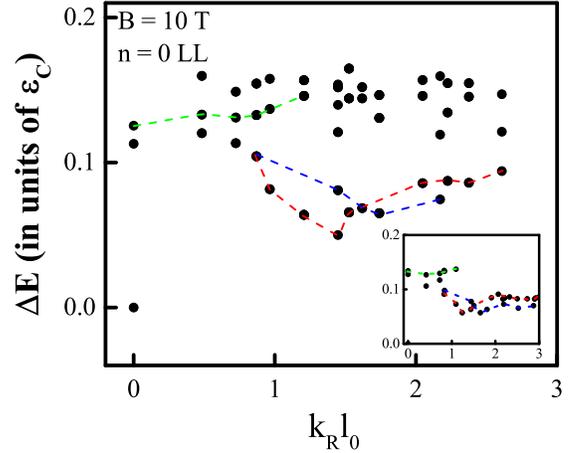}
\caption{\label{fig:DE_depk} The energy spectrum of the six-electron FQHE system in phosphorene 
with $\nu=\frac13$ in the lowest LL of the conduction band. The red and blue dash lines show 
schematically the two branches of the magnetoroton mode. Green dash line shows the second mode 
described in the text. Inset: the magnetoroton mode and the second mode for $N^{}_e=7$.
}\end{figure}  

Figure~\ref{fig:DE_depk} shows the energy spectrum of the six-electron FQHE system in the lowest LL of 
the conduction band for phosphorene with filling factor $\nu=1/3$. The inset shows 
the low energy collective modes for a seven electron system. As in conventional semiconductor 
systems, the magnetoroton mode can be clearly identified. The effect of anisotropy of the energy bands 
of phosphorene is that the magnetoroton mode is split into two branches with two minima. These are 
schematically shown in the figure with red and blue dashed lines. These kind of branches were 
actually noted previously in the context of adding a band mass tensor (Galilean metric) to the 
Hamiltonian in order to investigate the correlation between external and intrinsic metrics of the 
system \cite{HaldaneMetric,Qiu,Yang}. The intrinsic metric was treated as a variational parameter 
which characterizes the shape of the correlation function of the Laughlin state and is obtained by 
finding the maximum overlap of the exact wave function and the family of Laughlin wave functions 
parametrized by the intrinsic metric. The splitting of the magnetoroton mode into two branches
is possible in graphene in the case of anisotropic electron-electron interaction 
\cite{ApalkovMetric}. The appearance of two branches can be attributed to the fact that in the 
anisotropic system the minimum of the magnetoroton mode appears in different $|k^{}_R|$ for different 
direction of the relative momentum $\mathbf{k^{}_R}$. From this consideration we can conclude that 
phosphorene presents a real anisotropic system where the ideas of correlation of the external metric 
(mass tensor in this case) and the intrinsic metric can be explored.  

\begin{figure}
\includegraphics[width=7.5cm]{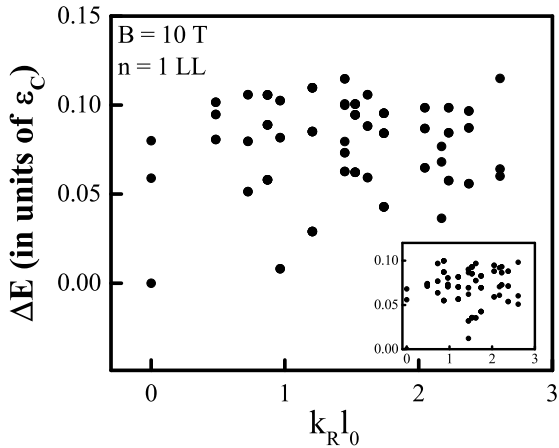}
\caption{\label{fig:DE_depk_SLL} The energy spectrum of the six-electron FQHE system with 
$\nu=\frac73$ in the second LL of conduction band for phosphorene. Inset: the same
spectrum but for a conventional semiconductor system (such as GaAs).}
\end{figure}

Interestingly, in Fig.~\ref{fig:DE_depk} we also observe a second mode 
(schematically shown by a green dashed line) for small wave vectors. It has positive dispersion 
and merges with the continuum for wave vectors $k^{}_\mathrm{R}\ell^{}_0>1$. This mode was observed 
previously in an experiment on the GaAs sample and was attributed to a two-roton state 
\cite{Hirjibehedin}. One possible theoretical explanation for the appearance of the second mode 
was that two types of shear forces exerted on the liquid state have different direction for different 
modes \cite{Tokatly}. While for isotropic systems these two modes should converge to the same point 
at $k^{}_\mathrm{R} \rightarrow0$ our studies indicate that for an anisotropic system these two modes 
are clearly separated for small wave vectors and do not converge to the same point as $k^{}_\mathrm{R}
\rightarrow0$. Therefore phosphorene can greatly simplify the identification of the second mode 
from the magnetoroton mode in resonant inelastic light scattering experiment. It should be noted 
that possibly anisotropy has played some role in the identification of the second mode even in the 
original experiment \cite{Hirjibehedin}. In that experiment the applied magnetic field is not perpendicular 
to the sample. The effect of the tilted magnetic field has been exhaustively studied 
in the experiments on FQHE and it was shown that tilting results in anisotropic transport properties 
of the sample \cite{XiaAnisotropy}. The effect of the tilted field on the FQHE states was also 
addressed theoretically by several authors \cite{Halonen,Papic} and it was found that inclusion 
of the tilted field is similar to adding an anisotropic metric in the system, although for the tilted 
magnetic field the dynamics is more complicated than simply the inclusion of the anisotropic mass. 
Therefore, due to the tilted field the system perhaps acquire some sort of 
anisotropy in the experiment \cite{Hirjibehedin}, which reveals the second mode.
Finally, we have performed similar studies for the highest LL in the valence band 
for $\nu=2/3$ and obtained very similar results. 

In Fig.~\ref{fig:DE_depk_SLL}, we present the energy spectrum of the six-electron FQHE system in 
the second LL of the conduction band for phosphorene with filling factor $\nu=7/3$. The same result 
for the conventional semiconductor system (such as GaAs) is shown as inset. It seems that the two 
branches of the magnetoroton mode and the second mode are still present. The gaps are however,
very small and the system seems to be approaching a soft mode, resulting in a phase transition
from  the incompressible to a compressible phase. Similar results are also found for the second LL 
in the valence band.

To summarize, we have studied the FQHE states in phosphorene in two LLs close to the band edges 
in the conduction and valence band. We have shown that the gaps for $\nu=1/3$ of 
electrons in the lowest LL of the conduction band and $\nu=2/3$ for holes in the highest LL of 
the valence band are slightly larger than for conventional semiconductor systems. Therefore the 
FQHE states should certainly be observable in phosphorene. The anisotropic band structure of 
phosphorene causes splitting of the magnetoroton mode into two branches with two minima. 
For small wave vectors, there is also a second mode with positive dispersion that is clearly separated 
from the magnetoroton mode. These features should be observable in inelastic light scattering 
experiments.

The work has been supported by the Canada Research Chairs Program of the Government
of Canada.


\begin{thebibliography}{99} 

\bibitem{graphene_book}
H. Aoki and M.S. Dresselhaus (Eds.), {\it Physics of Graphene}
(Springer, New York 2014).

\bibitem{abergeletal}
D.S.L. Abergel, V. Apalkov, J. Berashevich, K. Ziegler, and
T. Chakraborty, Adv. Phys. {\bf 59}, 261 (2010).

\bibitem{silicene}
P. De Padova, C. Ottaviani, C. Quaresima, B. Olivieri, P. Imperatori, E. Salomon, T. Angot,
L. Quagliano, C. Romano, A. Vona, M. Muniz-Miranda, A. Generosi, B. Paci and G. Le Lay,
2D Materials {\bf 1}, 021003 (2014); J. Sone, T. Yamagami, Y. Aoki, K. Nakatsuji and H. Hirayama,
New J. Phys. {\bf 16}, 095004 (2014).

\bibitem{germanene}
M.E. D\'avila, L. Xian, S. Cahangirov, A. Rubio and G. Le Lay, New J. Phys. {\bf 16}, 095002 (2014).

\bibitem{MoS2}
Q.H. Wang,  K. Kalantar-Zadeh, A. Kis, J. N. Coleman and M. S. Strano, Nature Nanotech. 7, 699–712 (2012).

\bibitem{LiNature}
L. Li, Y. Yu, G.J. Ye, Q. Ge, X. Ou, H. Wu, D. Feng, X.H. Chen and Y. Zhang, Nature Nanotech. {\bf 9}, 
372 (2014).

\bibitem{LiuAcsNano}
H. Liu, A.T. Neal, Z. Zhu, Z. Luo, X. Xu, D. Tom\'anek and P.D. Ye, ACS Nano {\bf 8}, 4033 (2014).

\bibitem{XiaNature}
F. Xia, H. Wang and Y. Jia, Nature Commun. {\bf 5}, 4458 (2014).

\bibitem{Kou}
L. Kou, C. Chen and S.C. Smith, J. Phys. Chem. Lett. {\bf 6}, 2794 (2015).

\bibitem{LiuReview}
H. Liu, Y. Du, Y. Deng and P.D. Ye, Chem. Soc. Rev. {\bf 44}, 2732 (2015).

\bibitem{LiHall1}
L. Li, G.J. Ye, V. Tran, R. Fei, G. Chen, H. Wang, J. Wang, K. Watanabe, T. Taniguchi, L. Yang,
X.H. Chen and Y. Zhang, Nature Nanotech. {\bf 10}, 608 (2015).

\bibitem{LiHall2}
L. Li, F. Yang, G.J. Ye, Z. Zhang, K. Watanabe, T. Taniguchi, Y. Wang, X.H. Chen and Y. Zhang,
arxiv:1504.07155 (2015).

\bibitem{Koenig}
S.P. Koenig, R.A. Doganov, H. Schmidt, A.H. Castro Neto and B. \"Ozyilmaz,
Appl. Phys. Lett. {\bf 104}, 103106 (2014).

\bibitem{Rodin}
A.S. Rodin, A. Carvalho and A.H. Castro Neto, Phys. Rev. Lett. {\bf 112}, 176801 (2014).

\bibitem{Low}
T. Low, A.S. Rodin, A. Carvalho, Y. Jiang, H. Wang, F. Xia and A.H. Castro Neto, 
Phys. Rev. B {\bf 90}, 075434 (2014).

\bibitem{Jiang}
Y. Jiang, R. Rold\'an, F. Guinea and T. Low, arxiv:1505.00175.

\bibitem{Rudenko}
A.N. Rudenko and M.I. Katsnelson, Phys Rev. B {\bf 89}, 201408 (2014).

\bibitem{Yuan}
S. Yuan, A.N. Rudenko and M.I. Katsnelson, Phys. Rev. B {\bf 91}, 115436 (2015).

\bibitem{Pereira}
J.M. Pereira Jr. and M.I. Katsnelson, arxiv:1504.02452.

\bibitem{Zhou}
X.Y. Zhou, R. Zhang, Y.L. Zou, D. Zhang, W.K. Lou, F. Cheng, G.H. Zhou, F. Zhai and K. Chang,
arxiv:1411.4275.

\bibitem{LiuNano}
Q. Liu, X. Zhang, L.B. Abdalla, A. Fazzio and A. Zunger, Nano Lett. {\bf 15}, 1222 (2015).

\bibitem{Ezawa}
M. Ezawa, New J. Phys. {\bf 16}, 115004 (2014).

\bibitem{Tahir}
M. Tahir, P. Vasilopoulos and F.M. Peeters, arxiv:1505.06780.

\bibitem{Cakir}
D. \c{C}akir, C. Sevik and F.M. Peeters, arxiv:1506.04707.

\bibitem{FQHEBook}
T. Chakraborty and P. Pietil\"ainen, {\it The Quantum Hall Effects} (Springer, New York, 1995);
{\it The Fractional Quantum Hall Effect} (Springer, New York, 1988).

\bibitem{FQHE_chapter}
T. Chakraborty and V. Apalkov, in \protect\cite{graphene_book} Ch. 8;
T. Chakraborty and V.M. Apalkov, Solid State Commun. {\bf 175}, 123 (2013).

\bibitem{ApalkovMonolayer}
V.M. Apalkov and T. Chakraborty, Phys. Rev. Lett. {\bf 97}, 126801 (2006).

\bibitem{ApalkovBilayer1}
V.M. Apalkov and T. Chakraborty, Phys. Rev. Lett. {\bf 105}, 036801 (2010).

\bibitem{ApalkovBilayer2}
V.M. Apalkov and T. Chakraborty, Phys. Rev. Lett. {\bf 107}, 186803 (2011).

\bibitem{FQHEGraphene}
X. Du, I. Skachko, F. Duerr, A. Luican and E.Y. Andrei, Nature {\bf 462}, 192 (2009);
K.I. Bolotin, F. Ghahari, M.D. Shulman, H.L. Stormer and P. Kim, Nature {\bf 462}, 196 (2009).

\bibitem{FQHEBilayer}
P. Maher, L. Wang, Y. Gao, C. Forsythe, T. Taniguchi, K. Watanabe, 
D. Abanin, Z. Papi\'c, P. Cadden-Zimansky, J. Hone, P. Kim and C.R. Dean, 
Science {\bf 345}, 61 (2014).

\bibitem{ApalkovBuckled}
V.M. Apalkov and T. Chakraborty, Phys. Rev. B {\bf 90}, 245108 (2014).  

\bibitem{Haldane}
F.D.M. Haldane, Phys. Rev. Lett. {\bf 55}, 2095 (1985).

\bibitem{PBCpapers}
T. Chakraborty, Surf. Sci. {\bf 229}, 16 (1990); Adv. Phys. {\bf 49}, 959 (2000);
T. Chakraborty and P. Pietil\"ainen, Phys. Rev. Lett. {\bf 76}, 4018 (1996);
T. Chakraborty and P. Pietil\"ainen, Phys. Rev. Lett. {\bf 83}, 5559 (1999);
T. Chakraborty and P. Pietil\"ainen, Phys. Rev. B {\bf 39}, 7971 (1989);
V.M. Apalkov, T. Chakraborty, P. Pietil\"ainen, and K. Niemel\"a, Phys. Rev. Lett.
{\bf 86}, 1311 (2001); T. Chakraborty, P. Pietil\"ainen, and F.C. Zhang, Phys. Rev.
Lett. {\bf 57}, 130 (1986); T. Chakraborty and F.C. Zhang, Phys. Rev. B {\bf 29},
7032 (R) (1984); F.C. Zhang and T. Chakraborty, Phys. Rev. B {\bf 30}, 7320 (R) (1984).

\bibitem{Peterson}
M.R. Peterson and C. Nayak, Phys. Rev. Lett. {\bf 113}, 086401 (2014).

\bibitem{HaldaneMetric}
F.D.M. Haldane, Phys. Rev. Lett. {\bf 107}, 116801 (2011).

\bibitem{Qiu}
R.-Z. Qiu, F.D.M. Haldane, X. Wan, K. Yang and S. Yi, Phys. Rev. B {\bf 85}, 
115308 (2012).

\bibitem{Yang}
B. Yang, Z. Papi\'c, E.H. Rezayi, R.N. Bhatt and F.D.M. Haldane, Phys. Rev. B 
{\bf 85}, 165318 (2012).

\bibitem{ApalkovMetric}
V.M. Apalkov and T. Chakraborty, Solid State Commun. {\bf 177}, 128 (2014). 

\bibitem{Hirjibehedin}
C.F. Hirjibehedin, I. Dujovne, A. Pinczuk, B.S. Dennis, L.N. Pfeiffer and K.W. West, 
Phys. Rev. Lett. {\bf 95}, 066803 (2005).

\bibitem{Tokatly}
I.V. Tokatly and G. Vignale, Phys. Rev. Lett. {\bf 98}, 026805 (2007).

\bibitem{XiaAnisotropy}
J. Xia, J.P. Eisenstein, L.N. Pfeiffer and K.W. West, Nat. Phys. {\bf 7}, 845 (2011).

\bibitem{Halonen}
T. Chakraborty and P. Pietil\"ainen, Phys. Rev. B {\bf 39}, 7971 (1989);
V. Halonen, P. Pietil\"ainen and T. Chakraborty, Phys. Rev. B {\bf 41}, 10202 (1990).

\bibitem{Papic}
Z. Papi\'c, Phys. Rev. B {\bf 87}, 245315 (2013).


\end{thebibliography}
\end{document}